\documentstyle[twoside,fleqn,espcrc2,epsf]{article}
\title
{Models of the Pseudogap State in Cuprates}

\author{M.V.Sadovskii
\address
{Institute for Electrophysics,\ Russian Academy of Sciences,\ Ural Branch,\\
Ekaterinburg,\ 620049,\ Russia}
\thanks{This work was supported in part
by the grant $N^{o}$ 96-02-16285 from the Russian Foundation for Basic
Research as well as by the grant No.20 of the State Program ``Statistical
Physics''.}}

\begin{document}

\begin{abstract}
We review a certain class of (``nearly'') exactly solvable models of 
electronic spectrum of two-dimensional systems with fluctuations 
of short range order of ``dielectric'' (e.g. antiferromagnetic) or
``superconducting'' type, leading to the formation of anisotropic
pseudogap state on certain parts of the Fermi surface. The models are based 
on recurrence procedure for one- and two-electron Green's functions which 
takes into account of all Feynman diagrams in perturbation series with the use
of the approximate Ansatz for higher-order terms in this series. 
These models can be applied to calculation of spectral density, density of
states and conductivity in the normal state, as well as to calculation of
some properties of superconducting state.
\end{abstract}

\maketitle

The model of ``nearly -- antiferromagnetic'' Fermi-liquid 
is based upon the picture of well developed fluctuations of AFM short range
order in a wide region of the phase diagram. In this model the
effective interaction of electrons with spin fluctuations is described via
dynamic spin susceptibility $\chi_{\bf q}(\omega)$, which is determined 
mainly from the fit to NMR experiments \cite{Sch}:  
\begin{equation} 
V_{eff}({\bf q},\omega)=g^2\chi_{\bf q}(\omega)\approx 
\frac{g^2\xi^2}{1+\xi^2({\bf q-Q})^2-i\frac{\omega}{\omega_{sf}}} 
\label{V} 
\end{equation} 
where $g$ is coupling constant, $\xi$--correlation length of spin 
fluctuations, ${\bf Q}=(\pi/a,\pi/a)$--vector of antiferromagnetic ordering 
in insulating phase, $\omega_{sf}$--characteristic frequency of spin
fluctuations, $a$--lattice spacing.
\begin{figure}
\epsfxsize=3cm
\epsfysize=3cm
\epsfbox{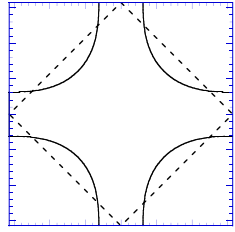}
\caption{``Hot spots'' model. Electronic states around
the intersection points of the Fermi surface with magnetic Brillouin zone
(shown by dashed lines) are strongly interacting with fluctuations of AFM
short range order.}
\end{figure}
As dynamic spin susceptibility $\chi_{\bf q}(\omega)$ has peaks at the wave
vectors around $(\pi/a,\pi/a)$ there appear ``two types'' of quasiparticles 
---``hot quasiparticles'' with momenta in the vicinity of ``hot spots'' on 
the Fermi surface and ``cold'' quasiparticles with momenta on the
other parts of the Fermi surface, e.g. around diagonals of the Brillouin 
zone $|p_x|=|p_y|$ \cite{Sch}. 

In the following we shall consider the case of high enough temperatures when
$\pi T \gg \omega_{sf}$ which corresponds to the region of ``weak pseudogap''
\cite{Sch}. In this case spin dynamics is irrelevant and we can
limit ourselves to static approximation.
\begin{figure}
\epsfxsize=6cm
\epsfysize=4cm
\epsfbox{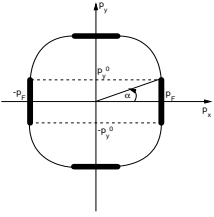}
\caption{Model of the Fermi surface for HTSC-cuprates. ``Hot patches''
are shown by thick lines of the width of $\xi^{-1}$).}
\end{figure}

We can greatly simplify all calculations if instead of (\ref{V}) we use
another form of model interaction:
\begin{equation}
V_{eff}({\bf q})=\Delta^2\frac{2\xi^{-1}}{\xi^{-2}+(q_x-Q_x)^2}
\frac{2\xi^{-1}}{\xi^{-2}+(q_y-Q_y)^2}
\label{Veff}
\end{equation}
where $\Delta$  is some phenomenological parameter,
determining the effective width of the pseudogap.
In fact (\ref{Veff}) is qualitatively
similar to the static limit of (\ref{V}) and differs from it very slightly 
in most important region of $|{\bf q-Q}|<\xi^{-1}$.

The spectrum of ``bare'' (free) quasiparticles can be taken in the form
\cite{Sch}:
\begin{equation}
\xi_{\bf p}=-2t(\cos p_xa+\cos p_ya)-4t^{'}\cos p_xa\cos p_ya
\label{spectr}
\end{equation}
where $t$--nearest neighbor transfer integral,   $t^{'}$--second nearest
neighbor transfer integral on the square lattice, $\mu$ is chemical potential.

Much simpler model of electronic spectrum
assumes that the Fermi surface of two -- dimensional system 
to have nesting (``hot'') patches of finite angular size $\alpha$
in $(0,\pi)$ and symmetric
directions in the Brillouin zone, as shown in Fig.2 \cite{PS}.
Similar Fermi surface was observed
in a number of ARPES experiments on cuprate superconductors.  
Here we assume that fluctuations interact only with electrons from
the ``hot'' (nesting) patches of the Fermi surface, and scattering vector
is either $Q_x=\pm 2p_F$,\ $Q_y=0$ or $Q_y=\pm 2p_F$,\ $Q_x=0$ for
incommensurate fluctuations, while
${\bf Q}=(\pi/a,\pi/a)$ for commensurate case.
It is easily seen that this scattering is in fact of one -- dimensional nature.
In this case non-trivial contributions of interaction with fluctuations appear 
only for electrons from ``hot'' patches, while electrons on ``cold'' parts of
the Fermi surface remain free.

These models can be solved exactly in the limit of infinite correlation length
$\xi\to\infty$, using methods developed in Refs. \cite{C1,C2}. For the case of
finite $\xi$ we can use an approximate Ansatz, proposed for one -- dimensional
case in Ref. \cite{C79} and further developed for two -- dimensional system in
Refs. \cite{Sch,KS}.
According to this Ansatz the contribution of an arbitrary diagram for electron 
self-energy of $N$-th order in interaction (\ref{Veff}) has the form:
\begin{equation}
\Sigma^{(N)}(\varepsilon_n{\bf p})=\Delta^{2N}\prod_{j=1}^{2N-1}
\frac{1}{i\varepsilon_n-\xi_{j}+in_jv_j\kappa}
\label{Ansatz}
\end{equation}
where for the ``hot spots'' model 
$\xi_j=\xi_{\bf p+Q}$ and $v_j=|v_{\bf p+Q}^{x}|+|v_{\bf p+Q}^{y}|$ for
odd $j$ and $\xi_j=\xi_{\bf p}$ and $v_{j}=|v_{\bf p}^x|+|v_{\bf p}^{y}|$
for even $j$ -- appropiate combinations of velocity projections, 
determined by the ``bare'' spectrum (\ref{spectr}).
For the ``hot patches'' model $\xi_j=(-1)^j\xi_{\bf p}$, $v_j=v_F$.
Here $n_j$ is the number of interaction lines, enveloping $j$-th
Green's function in a given diagram.
In this case any diagram with intersecting interaction lines is actually
equal to some diagram of the same order with noncrossing interaction lines.
Thus in fact we can consider only diagrams with nonintersecting interaction
lines, taking into account diagrams with intersecting lines introducing
additional combinatorial factors into interaction vertices. This method was
used for one-dimensional model of the pseudogap state in Refs.
\cite{C79,S91}.

As a result we obtain the following recursion relation for one-electron
Green's function (continuous fraction representation) \cite{C79,S91,KS}:
\begin{equation}
G^{-1}(\varepsilon_n\xi_{\bf p})=G_{0}^{-1}(\varepsilon_n\xi_{\bf p})-
\Sigma_{1}(\varepsilon_n\xi_{\bf p})
\label{G}
\end{equation}
\begin{equation}
\Sigma_{k}(\varepsilon_n\xi_{\bf p})=\Delta^2\frac{v(k)}
{i\varepsilon_n-\xi_k+ikv_k\kappa-\Sigma_{k+1}(\varepsilon_n\xi_{\bf p})}
\label{rec}
\end{equation}
Combinatorial factor:
\begin{equation}
v(k)=k
\label{vcomm}
\end{equation}
corresponds to the case of commensurate fluctuations with
${\bf Q}=(\pi/a,\pi/a)$ \cite{C79}. 
For incommensurate case \cite{C79}:
\begin{equation}
v(k)=\left\{\begin{array}{cc}
\frac{k+1}{2} & \mbox{for odd $k$} \\
\frac{k}{2} & \mbox{for even $k$}
\end{array} \right.
\label{vincomm}
\end{equation}
In Ref. \cite{Sch} a spin-structure of effective interaction within
the model of ``nearly antiferromagnetic'' Fermi-liquid was taken into 
account (spin-fermion model). This leads to more
complicated combinatorics of diagrams. 
Spin-conserving part of the interaction
gives formally commensurate combinatorics, while spin-flip scattering is
described by diagrams with combinatorics of incommensurate type.  
In this case \cite{Sch}:  
\begin{equation} 
v(k)=\left\{\begin{array}{cc}
\frac{k+2}{3} & \mbox{for odd $k$} \\
\frac{k}{3} & \mbox{for even $k$}
\end{array} \right.
\label{vspin}
\end{equation}
\begin{figure}
\epsfxsize=6cm
\epsfysize=6cm
\epsfbox{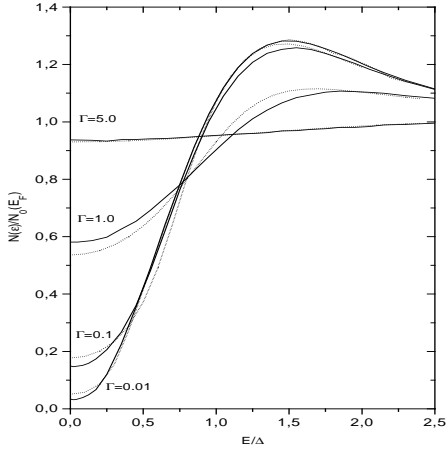}
\caption{Density of states in one -- dimensional version of the model
for incommensurate case and different values of $\Gamma=v_F\kappa/\Delta$.
Full lines -- exact numerical diagonalization,
dotted lines -- our Ansatz.}
\end{figure}
\begin{figure}
\epsfxsize=6cm
\epsfysize=6cm
\epsfbox{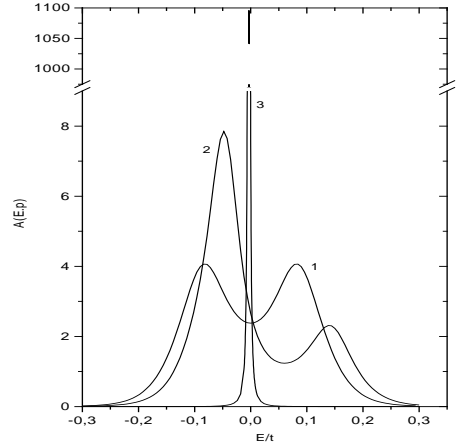}
\caption{
Energy dependencies of spectral density for $\kappa a=0.01$,
$t'/t=-0.4$, $\mu/t=-1.3$. Incommensurate case.
(1)---at the ``hot spot'' $p_xa/\pi =0.142$, $p_ya/\pi= 0.857$.
(2)---close to the ``hot spot''  
$p_xa/\pi =0.145$, $p_ya/\pi =0.843$.
(3)---far from the ``hot spot''  
$p_xa/\pi =p_ya/\pi =0.375$.}
\end{figure}

This approach was generalized for calculation of two -- electron Green's
function in Ref. \cite{S91}, where we formulated some recursion relations
for the vertex part, describing electronic response to an external 
electromagnetic field, allowing calculations of conductivity.

The Ansatz of (\ref{Ansatz}) is in fact not precisely exact \cite{Tch}, however
the analysis of Ref. \cite{KS}, shows that it is quantitatively good for most
interesting cases. These conclusions were confirmed in recent papers
\cite{Kopz,MM}, where the one -- dimensional version of our model was
solved by exact numerical diagonalization. In Fig.3 we show the comparison of
results obtained for the density of states in case of incommensurate
fluctuations by exact numerics \cite{Kopz}
and via our recursion relations. We can see an
extremely good correspondence, sufficient for any practical purposes.
In case of commensurate fluctuations in one -- dimension the Ansatz of
(\ref{Ansatz}) misses certain Dyson type singularity of the density of states,
appearing at the center of the pseudogap \cite{Kopz,MM}, 
but this is apparently absent in two -- dimensional case.

Consider one-electron spectral density:
\begin{equation}
A(E{\bf p})=-\frac{1}{\pi}Im G^R(E{\bf p})
\label{spd}
\end{equation}
where $G^R(E{\bf p})$ is retarded Green's function, obtained by the usual
analytic continuation of (\ref{G}) to the real axis of energy $E$.
In Fig.4 we show typical energy dependencies of $A(E{\bf p})$ obtained for 
the ``hot spots'' model. More detailed results can be found in 
Refs. \cite{Sch,KS}. Similar non Fermi -- liquid like behavior of the spectral 
density is easily obtained on ``hot patches'' of the Fermi surface shown in 
Fig.2 (Cf. Ref. \cite{S91}).

Let us stress 
that our solution (\ref{rec}) is exact both for $\xi\to\infty$ and
$\xi\to 0$, while in the region of finite $\xi$ it provides apparently very good
interpolation.

The one-electron density of states:
\begin{equation}
N(E)=\sum_{\bf p}A(E,{\bf p})=-\frac{1}{\pi}\sum_{\bf p}ImG^R(E{\bf p})
\label{dos}
\end{equation}
is determined by the integral of spectral density $A(E{\bf p})$ over 
the Brillouin zone. Detailed results for the ``hot spots'' model
were obtained in Ref. \cite{KS}, demonstrating smeared pseudogap with
weak dependence on the correlation length $\xi$. For ``hot patches'' model
this pseudogap is more pronounced, depending on the size of these patches. 

``Hot patches'' model was applied by us to calculations of optical conductivity
\cite{C99op} and also (for the simplest case of $\xi\to\infty$) to the study of
the main superconducting properties \cite{PS,C99}.

Pseudogap phenomena can also be  explained using the idea of
fluctuation Cooper pairing at temperatures higher than superconducting
transition temperature $T_c$.
Anticipating the possibility of both $s$-wave
and $d$-wave pairing, we introduce the
pairing interaction of the simplest (separable) form:
\begin{equation}
V({\bf p,p'})=-Ve(\phi)e(\phi')
\label{Vsc}
\end{equation}
where $\phi$ is polar angle determining the direction of electronic momentum
${\bf p}$ in the plane, while for $e(\phi)$ we assume model dependence:
\begin{equation}
e(\phi)=\left\{\begin{array}{cc}
1 & \mbox{$s$-wave pairing}\\
\sqrt{2}\cos(2\phi) & \mbox{$d$-wave pairing}
\end{array} \right.
\label{e}
\end{equation}
Analogously to transition from (\ref{V}) to (\ref{Veff}) we introduce
the model interaction (static fluctuation propagator of Cooper pairs):  
\begin{equation} 
V_{eff}({\bf q})=-\Delta^2 e^2(\phi)\frac{2\xi^{-1}}{\xi^{-2}+q_x^2} 
\frac{2\xi^{-1}}{\xi^{-2}+q_y^2}
\label{Veffsc}
\end{equation}
where $\Delta$ determines the width of superconducting pseudogap.
We can see that mathematically 
this problem is practically the same as 
the ``hot spot'' model, but always with combinatorics of incommensurate case
\cite{KS}.
Then we again obtain the recurrence relation for the Green's function of
the type of (\ref{rec}):  
\begin{eqnarray} 
\Sigma_{k}(\varepsilon_n\xi_{\bf p})=\nonumber\\
\frac{\Delta^2e^2(\phi)v(k)} {i\varepsilon_n-(-1)^k\xi_{\bf 
p}+ik(|v_x|+|v_y|)\kappa- \Sigma_{k+1}(\varepsilon_n\xi_{\bf p})} 
\label{recsc} 
\end{eqnarray} 
where $v_x=v_F\cos\phi$, $v_y=v_F\sin\phi$, $\kappa=\xi^{-1}$, $\varepsilon_n>0$
and $v(k)$ was defined in (\ref{vincomm}).

Energy dependencies of the spectral density 
$A(E{\bf p})$ for one-particle Green's function (\ref{spd}), can be
calculated from
(\ref{recsc}) for different values of polar angle $\phi$, determining the
direction of electronic momentum in the plane,
for the case of fluctuations of $d$-wave pairing. Calculations show \cite{KS}
that in
the vicinity of the point $(\pi/a,0)$ in Brillouin zone this spectral density
is non Fermi-liquid like (pseudogap behavior). 
As vector ${\bf p}$ rotates in the
direction of the zone diagonal the two peak structure gradually disappears
and spectral density transforms to the typical Fermi-liquid like with a
single peak, which narrows as $\phi$ approaches $\pi/4$. Analogous
transformation of the spectral density takes place as correlation length
$\xi$ becomes smaller. 
In case of fluctuation pairing of $s$-wave type the pseudogap appears
isotropically on the whole Fermi-surface.


\begin{thebibliography}{99}

\bibitem{Sch} J.Schmalian,~D.Pines,~B.Stojkovic. Phys. Rev. Lett. {\bf 80},
3839(1998),\ Phys. Rev. {\bf B60}, 667 (1999)
\bibitem{C1} M.V.Sadovskii. Sov.Phys.-JETP {\bf 39}, 845 (1974)
\bibitem{C2} M.V.Sadovskii. 
Sov.Phys.-Solid State {\bf 16}, 1632 (1974)
\bibitem{C79}M.V.Sadovskii. Sov.Phys.-JETP {\bf 50}, 989 (1979)
\bibitem{S91}M.V.Sadovskii, A.A. Timofeev. J.Moscow Phys.Soc. {\bf 1}, 391(1991)
\bibitem{PS}A.I.Posazhennikova, M.V.Sadovskii. JETP {\bf 88}, 347 (1999)
\bibitem{KS}E.Z.Kuchinskii, M.V.Sadovskii. JETP {\bf 88}, 347 (1999)
\bibitem{Tch}O.Tchernyshyov. Phys. Rev. {\bf B59}, 1358 (1999)
\bibitem{Kopz}L.Bartosch,~P.Kopietz. Preprint cond-mat /9908065
\bibitem{MM}A.J.Millis,~H.Monien. Preprint cond-mat /9907223
\bibitem{C99op}M.V.Sadovskii. JETP Lett. {\bf 69}, 447 (1999) and these
proceedings
\bibitem{C99}E.Z.Kuchinskii, M.V.Sadovskii. JETP (in press) and these
proceedings

\end{thebibliography}
\end{document}